\documentclass[12pt,letterpaper]{article}
\usepackage{jheppub}
\usepackage{graphicx} 
\input{epsf}
\usepackage{subfigure}

\usepackage{verbatim}

\title{AdS black holes, the bulk-boundary dictionary, and smearing functions}

\author[a]{Stefan Leichenauer }
\affiliation[a]{California Institute of Technology, Pasadena, CA 91125, U.S.A.}
\emailAdd{sleichen@theory.caltech.edu}

\author[b]{and Vladimir Rosenhaus }
\affiliation[b]{Center for Theoretical Physics and Department of Physics,\\
 University of California, Berkeley, CA 94720, U.S.A.}
\emailAdd{vladr@berkeley.edu}

\abstract{In Lorentzian AdS/CFT there exists a mapping between local bulk operators and nonlocal CFT operators. In global AdS this mapping can be found through use of bulk equations of motion and allows the nonlocal CFT operator to be expressed as a local operator smeared over a range of positions and times. We argue that such a construction is not possible if there are bulk normal modes with exponentially small near boundary imprint. We show that the AdS-Schwarzschild background is such a case, with the horizon introducing modes with angular momentum much larger than frequency, causing them to be trapped by the centrifugal barrier. More generally, we argue that any barrier in the radial effective potential which prevents null geodesics from reaching the boundary will lead to modes with vanishingly small near boundary imprint, thereby obstructing the existence of a smearing function. While one may have thought the bulk-boundary dictionary for low curvature regions, such as the exterior of a black hole, should be as in empty AdS, our results demonstrate otherwise. }

\begin{document}
\maketitle

\section{Introduction}

The Lorentzian AdS/CFT \cite{Mal97, Wit98a, GubKle98} dictionary in extrapolate form gives a simple relation between a bulk operator close to the boundary, and a boundary operator. If $\Phi(B)$ is a bulk operator, where $B$ denotes a bulk coordinate $B = (r, t, \Omega)$, and $b$ is a boundary coordinate $b = (t, \Omega)$, then \cite{BanDoug98, Gid99, Har11}
\begin{equation} \label{eq:DIC}
\lim_{r \rightarrow \infty} r^{\Delta} \Phi(B) = O(b).
\end{equation}
This relates a local bulk operator at large $r$ to a local boundary operator. But what is the CFT dual of $\Phi(B)$ at finite $r$? A natural proposal is 
\begin{equation} \label{eq:Kabat}
\Phi(B) = \int{db' ~K(B|b') O(b')} + \mathcal{O}(1/N),
\end{equation}
where $K(B|b')$ is some smearing function. 

There is no reason a dictionary as simple as (\ref{eq:Kabat}) has to be true.  Our goal in this paper will be to make progress on establishing when a mapping like (\ref{eq:Kabat}) is and isn't possible. In pure global AdS, the smearing function $K(B|b')$ was found in \cite{HamKab06}. We will show a smearing function as in (\ref{eq:Kabat}) does not exist in AdS-black hole backgrounds, for \textit{any} bulk point $B$. 

Eq.~\ref{eq:Kabat} is in some ways an extraordinary statement. It allows us to express an $n$-point function of bulk operators $\Phi$ in terms of smeared $n$-point functions of boundary operators $O$. Thus it says that the entire bulk state is encoded in terms of operators $O$. Yet, the operators $O$ are special: they are related via (\ref{eq:DIC}) to the large $r$ limit of local bulk operators. Most field theory operators, for instance Wilson loops, are not of this form.\footnote{One should keep in mind that in (\ref{eq:Kabat}) we are smearing on the boundary over both space and time; if one were to use the CFT Hamiltonian to evolve the right side of (\ref{eq:Kabat}) to a single time then one would generate an operator with Wilson loops. Nevertheless, the ability to avoid Wilson loops if one is allowed to compute correlation functions of the $O$ for different times is in itself nontrivial. }

Having the precise form of the smearing function $K(B|b')$ is an important component of the AdS/CFT dictionary. For any bulk point $B$, $K(B|b')$ will presumably have most of its support on some subregion of the boundary. Thus, $K(B|b')$ would tell us, independent of the state, which subregion  \cite{BouLei12, CzeKar12, HubRan12} of the global AdS boundary is ``responsible'' for a bulk point $B$.\footnote{We should note that in general $K(B|b')$ will not be unique.}

As a result of (\ref{eq:DIC}), Eq.~\ref{eq:Kabat} has a purely bulk interpretation. It states that a bulk operator at point $B$ can be expressed in terms of smeared bulk operators at large radius. Alternatively, in the Schr\"odinger picture it states that the bulk wavefunction restricted to a large but fixed $r=R$ and provided for some sufficient time extent, completely encodes the bulk state for all $r<R$. It is not obvious if this is a true or false statement. If a particle is sitting at the center of AdS, its wavefunction at large $r$ will be small but nonvanishing; perhaps that is enough to determine the wavefunction everywhere? Or perhaps there are some states for which the wavefunction has vanishingly  small support at $r=R$, making (\ref{eq:Kabat}) impossible?

While (\ref{eq:Kabat}) is an operator statement, determining $K(B|b')$ is a classical field theory problem: Given $\varphi(b') \equiv \Phi(r=R, t', \Omega')$, how does one reconstruct $\Phi(B)$? This is a nonstandard boundary value problem, with data being specified on a timelike surface. However, having a smearing function is a more stringent requirement than simply having an algorithm for deterimining $\Phi(B)$ for any given $\varphi(b')$. For instance, it may be the case that for any particular bulk solution, even if $\varphi(b')$ is extremely small, one can pick an appropriate resolution so as to see it and reconstruct $\Phi(B)$. However, it could be that no matter how good a resolution one picks, there always exist field configurations having a near boundary imprint $\varphi(b')$ that is below the resolution scale. In such a case there wouldn't be a smearing function; for a smearing function implies a state-independent way of reconstructing. In a sense one has to pick the resolution beforehand without knowledge of which field configurations will be under consideration. As a CFT statement, the absence of a smearing function means that certain aspects of the bulk are not well encoded in the smeared CFT operators $O(b')$, but rather in the more general Wilson loops. 

Constructing a smearing function is straightforward in static, spherically symmetric spacetimes. One solves the bulk equations of motion through a mode decomposition: $\Phi_{\omega l m}(r ,t, \Omega) = \phi_{\omega l}(r) Y_{l m}(\Omega) e^{-i \omega t}$. The bulk is reconstructed mode by mode, using the boundary imprint to extract the coefficient of each mode. In some cases this can be used to construct a smearing function. However in other cases, for reasons discussed above and which we will make precise in Sec.~\ref{sec:2}, the candidate smearing function is a divergent sum. Pure AdS falls into the first category, while AdS-black holes are in the second. 

In AdS, just like in flat space, at small $r$ there is a centrifugal barrier which reflects the modes. However, black holes have the property that at a finite distance from the horizon the centrifugal barrier peaks and the potential dies off as the horizon is approached. Unlike in pure AdS, modes with $\omega \ll l$ become admissible, and are trapped behind the centrifugal barrier. As $l$ is increased with $\omega$ kept constant, the barrier grows, and the imprint of the modes at large $r$ decays exponentially in $l$. 

In Sec.~\ref{sec:2} we review the mode sum approach to obtaining a smearing function.  In Sec.~\ref{sec:3} we rewrite the Klein-Gordon equation for a scalar field in a static, spherically symmetric background as a Schr\"odinger equation. For large $l$, the potential  has roughly two competing terms: the centrifugal barrier and the AdS barrier $ \sim r^2$. For any radius $r$, no matter how large, there is an $l$ sufficiently high so that the centrifugal barrier dominates. In Sec.~\ref{sec:4} we use WKB to show that in AdS-black hole backgrounds this effect gives rise to the exponential behavior in $l$ for the modes. 

It may seem surprising that our ability to describe the bulk at large $r$, where the metric is well approximated by pure AdS, could be affected by the presence of a small black hole deep in the bulk. In Sec.~\ref{sec:4.1} we show that while it is true the behavior of the modes near the boundary is always well approximated by Bessel functions $r^{-d/2}J_{\nu}(\sqrt{\omega^2 - l(l+d-2)}/r)$, the relation between $\omega$ and $l$ depends on the entire bulk geometry. In pure AdS, $\omega$ is quantized as $\omega_n = 2n +l +\Delta$, while in the AdS-black hole $\omega$ is continuous and independent of $l$. As a result, $\omega \ll l$ is allowed for AdS-black holes, leading the Bessel function to have imaginary argument and correspondingly exhibit exponential growth. 

In Sec.~\ref{sec:5} we take the first steps towards generally establishing for which asymptotically AdS spacetimes a smearing function exists. In Sec.~\ref{sec:5.1} we consider a  static, spherically symmetric spacetime and argue that only the behavior of the high $l$ modes is relevant for this question. We find in this limit the potential in the Schr\"odinger-like equation describing the modes simplifies significantly. We find that any barrier in the large $l$ potential leads to exponential behavior in $l$ of the modes and prevents a smearing function. Here the appropriate limit leading to an exponentially suppressed tail involves sending $\omega$ to infinity as well sending $l$ to infinity, while keeping the ratio $\omega/l$ constant. Thus, we will find that even a small, dense star in AdS can prevent a smearing function from existing for some bulk points. However, unlike the black hole, we are not necessarily prevented from constructing a smearing function at large $r$ in general. In Sec.~\ref{sec:5.2} we consider general spacetimes and examine the possibility of the existence of trapped null geodesics (geodesics with neither endpoint on the boundary) as a proxy for the smearing function not existing. We find that in static spherically symmetric spaces the null geodesic equation is that of a classical point particle moving in a potential identical to the one found in Sec.~\ref{sec:5.1} as being relevant for the smearing function question. Therefore a smearing function does not exist if there are trapped null geodesics. 

\section{Smearing functions} \label{sec:2}

We work in Lorentzian AdS$_{d+1}$/CFT$_d$ with fixed boundary Hamiltonian, and correspondingly all nonnormalizable bulk modes turned off. We let $B$ denote a bulk coordinate, $B = (r, t, \Omega)$, and $b$ a boundary coordinate $b = (t, \Omega)$. If we consider a scalar field $\Phi(B)$ in the bulk, then excited states are obtained by acting with  $\Phi(B)$ on the vacuum. As the boundary is approached, $\Phi(r \rightarrow \infty)$ will decay to 0. However, we can extract the leading term $\varphi$ in the decay, $\Phi(B) \rightarrow \varphi(b)/r^{\Delta}$, where the conformal dimension $\Delta = d/2 + \sqrt{d^2/4 +m^2}$. The extrapolate version of the AdS/CFT dictionary instructs us to identify $\varphi$ with a local boundary operator: $\varphi(b) \leftrightarrow O(b)$. 

As a result we can construct a relation between $\Phi(B)$ and the CFT operators $O(b)$ by relating the tail $\varphi$ of $\Phi$ at the boundary to $\Phi(B)$ through use of bulk equations of motion. This is a nonstandard boundary value problem where the data is specified on a timelike surface. Unlike usual time evolution where the field at point is determined by the field in the causal past of that point, here we have little intuition about which portion of the boundary is needed to determine $\Phi(B)$.

\subsubsection*{Smearing function as a mode sum }
In the limit of infinite $N$, the bulk field operator $\Phi (B)$ obeys the free wave equation and its reconstruction from boundary data can be implemented through Fourier expansion. Letting $\Phi_k(B)$ be the orthogonal solutions to the Klein-Gordon equation (where $k$ is a collective index), we do a mode expansion of $\Phi(B)$ in terms of creation and annihilation operators $a_k$, 
\begin{equation} 
\Phi(B) = \int{dk  ~a_k \Phi_k(B) + h.c.}
\end{equation}
Taking $B$ to the boundary and letting $\varphi_k = \Phi_k r^{\Delta}$ gives 
\begin{equation} \label{eq:O}
O(b) = \int{dk ~a_k \varphi_k(b) + h.c.}
\end{equation}
In some cases the boundary mode functions $\varphi_k(b)$ are orthogonal. If they are  we can invert (\ref{eq:O})
\begin{equation} \label{eq:a}
a_k = \int{db ~O(b) \varphi_k^{*}(b)} ,
\end{equation}
where we have with hindsight chosen to normalize the modes $\Phi_k$ so that $\varphi_k$ are orthonormal.\footnote{One is of course free to choose any normalization for the $\Phi_k$; however if the $\varphi_k$ are not orthonormal, (\ref{eq:K}) will get modified by the appropriate factor.}
Inserting (\ref{eq:a}) into (\ref{eq:O}) gives

\begin{equation} \label{eq:Phi2}
\Phi(B) = \int{dk ~\left[\int{db' ~\varphi_{k}^{*}(b') O(b')}\right] \Phi_k(B)+ h.c} .
\end{equation}
Exchanging the integrals over $k$ and $b$ gives
\begin{equation} \label{eq:Smearing}
\Phi(B) = \int{db' ~K(B|b') O(b')} ,
\end{equation}
where 
\begin{equation} \label{eq:K}
K(B|b') = \int{dk ~\Phi_k(B) \varphi_k^*(b') +h.c.}
\end{equation}

\subsubsection*{Potential divergences of the smearing function}
Eq. \ref{eq:K} is the equation for a smearing function and will be the focus of the rest of the paper. In all the cases we will consider, (\ref{eq:a}) will exist, but the integral in (\ref{eq:K}) may or may not converge. In the limit of infinite $N$ the question of the existence of a smearing function in some background can therefore be equivalently stated as the question of convergence of the integral in (\ref{eq:K}).\footnote{There is a potential loophole. The smearing function could exist without the integral (\ref{eq:K}) converging. If this were the case, the smearing function would have to be a function whose Fourier transform is not well-defined.}
In cases when a smearing function exists in the $N=\infty$ limit,  one can then include corrections to (\ref{eq:Smearing}) perturbatively in $1/N$ \cite{Kab11,HeeMar11}. We will only be concerned with the smearing function at infinite $N$.

The bulk modes $\Phi_k$ that appear in (\ref{eq:K}) need to be normalized so that the boundary modes $\varphi_k$ they asymptote to are orthonormal.  This is at the heart of the problem of constructing smearing functions. When we decompose some bulk solution $\Phi(B)$ in terms of modes, we generally don't expect each mode to be weighted equally. Rather, there are some modes which may have small coefficients. However, the smearing function is state independent and has no way of knowing which modes will get small weight. 

When working in a general background, not all modes are equal. Some modes may need to pass through enormous barriers in the potential on their way to the boundary and consequently suffer a huge damping. All the modes $\Phi_k$ are normalized so that the $\varphi_k \equiv \Phi_k r^{\Delta}$ they asymptote to on the boundary are orthonormal. As a result, modes that had to pass through a large barrier will be extremely large at small $r$.   For any particular solution this wouldn't bother us, as these modes would have a small expectation value of $a_k$. As a result, (\ref{eq:Phi2}) would converge. However, without having the small $a_k$ to dampen the modes at small $r$, the integral (\ref{eq:K}) appearing in (\ref{eq:Smearing}) might diverge. As we will see later, this is precisely what happens in AdS-Rindler and in AdS-Schwarzschild. 

\subsubsection*{Pure AdS smearing function}
\begin{figure}[tbp]
\centering
\subfigure[]{
	\includegraphics[width=2in]{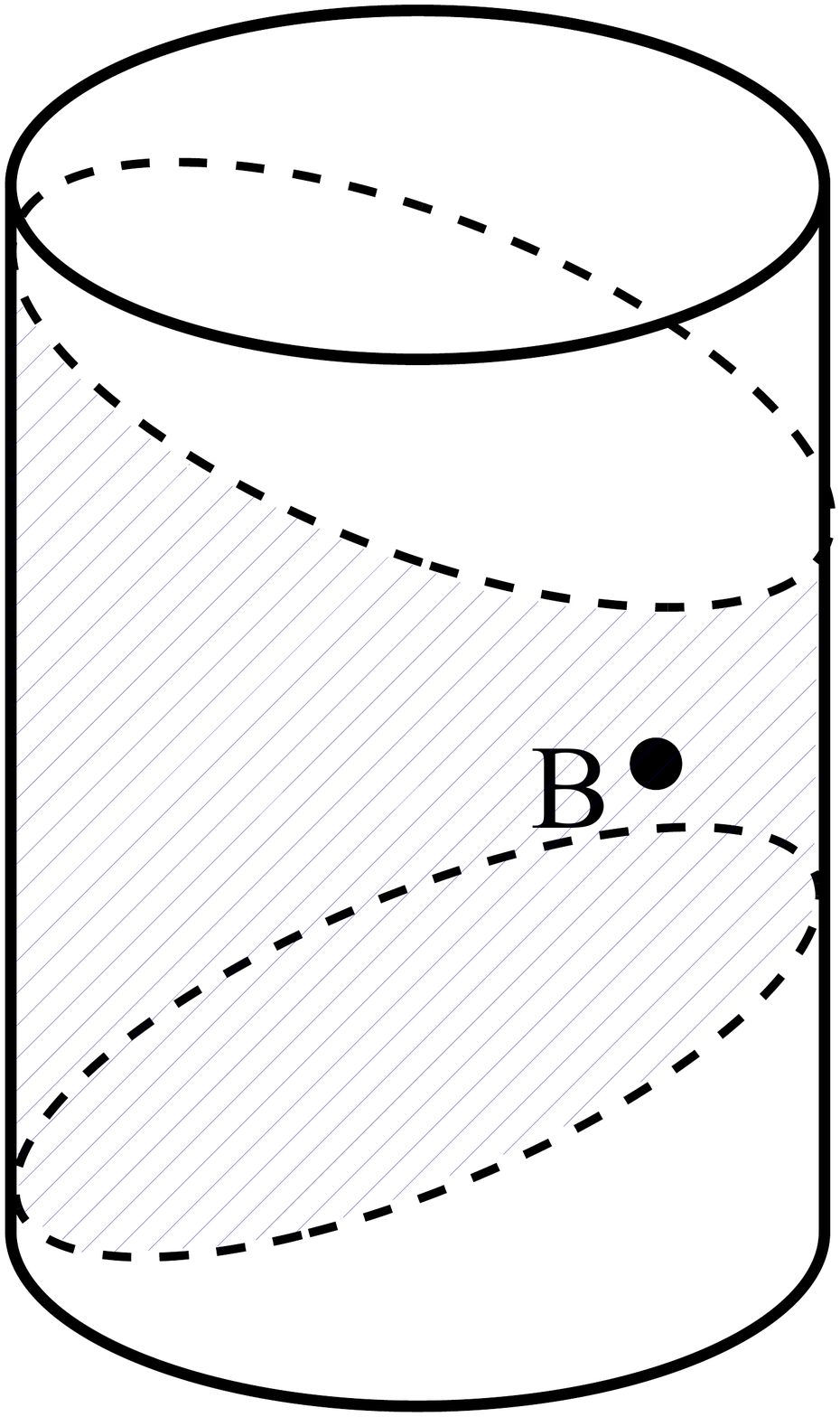}
	}
	\hspace{.2in}
		\subfigure[]{
	\includegraphics[width=2in]{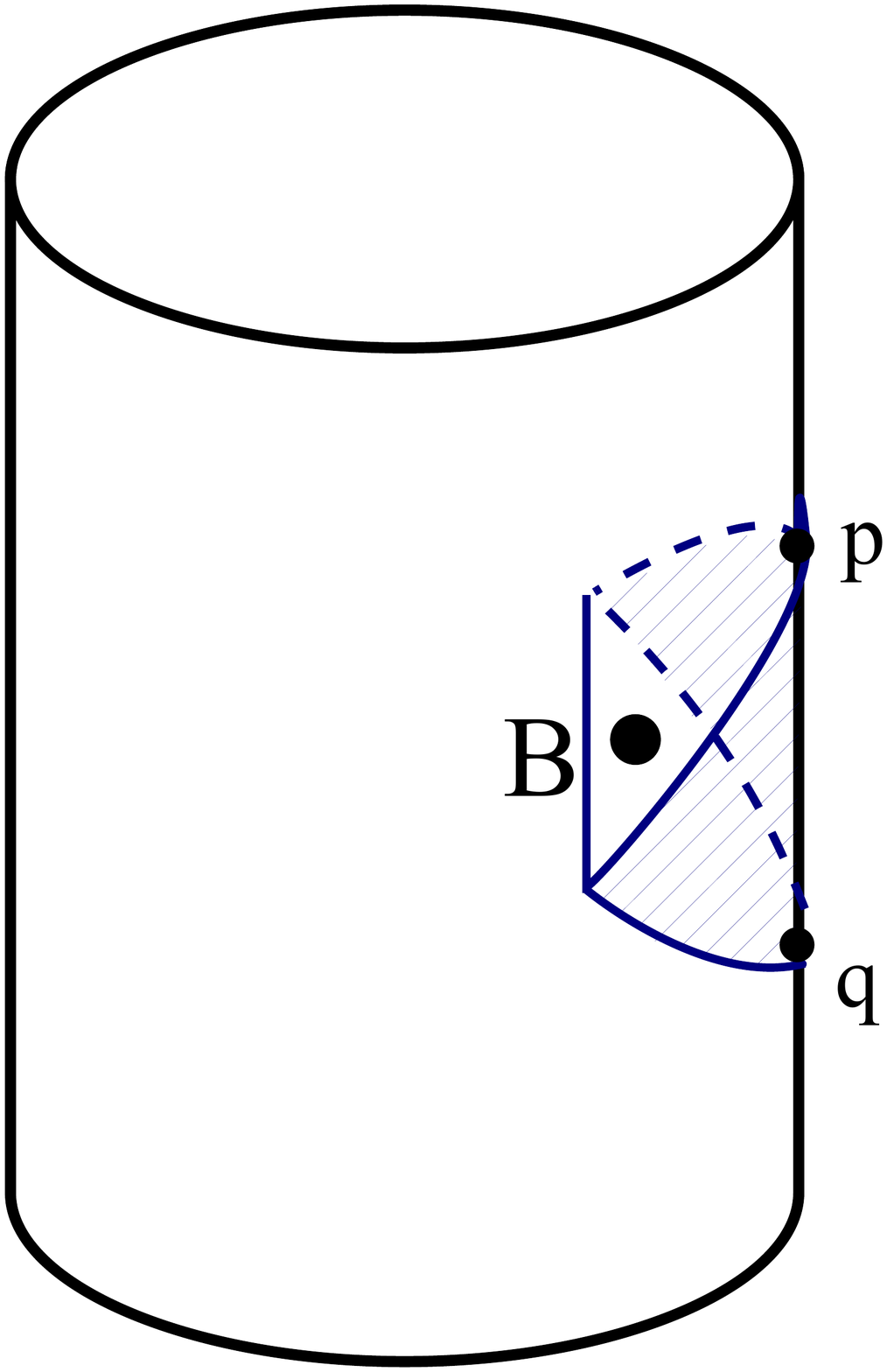}}
\caption{ To construct the bulk operator $\Phi(B)$, the CFT operator $O(b')$ is smeared with the smearing function $K(B|b')$ as indicated in (\ref{eq:Smearing}). (a) The support of the pure AdS smearing function $K(B|b')$ is all boundary points $b'$ spacelike separated from $B$ (hatched region). (b) Had the AdS-Rindler smearing function existed, it would have only made use of the boundary region that overlaps with $J^+(q)\cap J^-(p)$ (the intersection of the causal future of $q$ and causal past of $p$), where $q$ and $p$ are chosen so that $J^+(q)\cap J^-(p)$ just barely contains $B$. Any changes outside this bulk region $J^+(q)\cap J^-(p)$ would have been manifestly irrelevant for computing $\Phi(B)$.}  \label{fig:Smearing}
\end{figure}

The metric for global AdS$_{d+1}$ can be written as
\begin{equation} \label{eq:global}
ds^2 = -(1+r^2)dt^2 + \frac{dr^2}{1+r^2} + r^2 d\Omega_{d-1}^2 .
\end{equation}
The smearing function $K(B|b')$ was constructed in \cite{HamKab06} (see also \cite{Bena:1999jv, HamKab05}).\footnote{To avoid any potential confusion, we note that in Lorentzian AdS/CFT the smearing function problem is distinctly different from the one Witten's bulk-boundary propagator \cite{Wit98a} addressed in Euclidean space. The Lorentzian version of Witten's bulk-boundary propagator is a bulk Green's function with one point taken to the boundary (as a result, unlike (\ref{eq:GSmear}) it does manifestly approach a delta function for coincident points). However, it is a smearing function for the nonnormalizable modes, which are dual to sources for the CFT; whereas we are interested in a smearing function for normalizable modes.}
Notably, it has support on boundary points $b'$ that are spacelike (or null) seperated from $B$ (shown in Fig.~\ref{fig:Smearing}). 
It takes a different form in even and odd dimensions, and is simpler when $d+1$ is even:
\begin{equation} \label{eq:GSmear}
K(B|b') = \left[\sqrt{1+r^2} \cos(t-t') - r \cos( \Omega - \Omega')\right]^{\Delta-d}  .
\end{equation}

The spacelike support of $K(B|b')$ gives it some peculiar features. If one uses $K(B|b')$ to construct $\Phi(B)$ through (\ref{eq:Smearing}) and considers the limit of taking $B$ to the boundary, it is not manifest that $\Phi(B) \rightarrow r^{-\Delta} O(b)$. In fact, UV/IR \cite{SusWit98} seems to suggest one should only need some compactly supported portion of the boundary to construct $\Phi(B)$ if $B$ is close to the boundary. However, the smearing function (\ref{eq:GSmear}) does not reflect this intuition. Indeed, the limit of (\ref{eq:GSmear}) when $B$ is close to the boundary, 
\begin{equation}
K(r\rightarrow \infty, t, \Omega \ | b') \rightarrow r^{\Delta -d } \left[ \cos(t-t') -\cos(\Omega - \Omega')\right]^{\Delta-d}
\end{equation}
is not at all peaked at small $t-t'$ and $\Omega-\Omega'$. 

\subsubsection*{AdS-Rindler smearing function}

Perhaps the smearing function (\ref{eq:GSmear}) is not optimal and uses more boundary data than actually necessary? The minimal possible boundary region (shown in Fig.~\ref{fig:Smearing}b) can be found by picking boundary points $p,q$ such that $B$ is just barely contained in the intersection of the causal future of $q$ with the causal past of $p$, $J^+(q)\cap J^-(p)$. The intersection of $J^+(q)\cap J^-(p)$ with the boundary yields the smallest boundary region allowed by causality~\cite{Bou09}. A convenient coordinate system to use which covers only this region is AdS-Rindler, which in $AdS_3$ takes the form
\begin{equation} \label{eq:AdSRindler}
ds^2 = -(r^2-1)dt^2 + \frac{dr^2}{r^2-1} + r^2 dx^2 .
\end{equation}
In Ref. \cite{HamKab06} construction of a smearing function of this form was attempted, but the procedure fails. The solution for the modes in terms of a hypergeometric function is~\cite{Kesk}
\begin{equation} \label{eq:hyp}
\Phi_{\omega k} (r, t, x) = r^{-\Delta} \left(\frac{r^2-1}{r^2}\right)^{-i \omega/2} F\left(\frac{\Delta - i \omega - i k}{2},\frac{\Delta - i \omega + i k}{2}, \Delta, \frac{1}{r^2}\right)   e^{i(kx -\omega t)} .
\end{equation}
For $k\gg\omega$ the modes have an exponential growth in $k$. As a result, the integral  (\ref{eq:K}) doesn't converge. Note that  although  AdS-Rindler asymptotes to the Poincare Patch at large $r$, modes with $k\gg \omega$ are forbidden in Poincare Patch but allowed in AdS-Rindler.

Had an AdS-Rindler smearing existed, it would have guaranteed a smearing function for points $B$ in the large $r$ region of any asymptotically AdS geometry. The field and metric at any point outside of $J^+(q)\cap J^-(p)$ would have been manifestly irrelevant. In the absence of an AdS-Rindler smearing function, all we have is the global smearing function. Since it makes use of the entire spacelike separated region from $B$, changes to the field anywhere in the bulk could potentially have an impact on reconstruction of $\Phi(B)$. While we wouldn't expect a small perturbation of the metric at the center of AdS to have a significant impact on the form of the smearing function, a black hole in the center is a major change to metric and the existence of a smearing function is no longer guaranteed.

Our goal will be to understand in which circumstances the smearing function does and doesn't exist; when (\ref{eq:K}) does and doesn't converge. In the following section we analyze the bulk modes in an AdS-Schwarzschild background. 

\section{Solving the wave equation} \label{sec:3}

In this section we rewrite the wave equation for a scalar field in the form of a Schr\"odinger equation, allowing us to easily analyze the solutions. 

We consider a scalar field $\Phi(B)$ in a static, spherically symmetric background,
\begin{equation} \label{eq:Static}
ds^2 = - f(r) dt^2 + \frac{dr^2}{h(r)} + r^2 d\Omega_{d-1}^2.
\end{equation}
To leading order in $1/N$ the field $\Phi$ satisfies the free wave equation
\begin{equation}
\frac{1}{\sqrt{g}}\partial_{\mu}(\sqrt{g}g^{\mu \nu} \partial_{\nu} \Phi) - m^2 \Phi =0 .
\end{equation}
Separating $\Phi$ as

\begin{equation} \label{eq:phi}
\Phi(r,t,\Omega) = \phi(r) Y(\Omega) e^{- i \omega t}
\end{equation}
gives for the radial field $\phi(r)$,
\begin{equation} \label{eq:rad1}
\frac{\omega^2}{f} \phi + \frac{1}{r^{d-1}}\sqrt{\frac{h}{f}}\partial_r (\sqrt{f h}r^{d-1} \partial_{r} \phi) - \frac{l(l+d-2)}{r^2} - m^2 \phi = 0.
\end{equation}
Letting $\phi(r) = u(r)/r^{\frac{d-1}{2}}$ and changing variables to a tortoise-like coordinate $dr_{*} = dr/\sqrt{f h}$ turns (\ref{eq:rad1}) into a Schr\"odinger-like equation
\begin{equation} \label{eq:Schro}
\frac{d^2 u}{dr_{*}^2} + (\omega^2 - V(r))u=0,
\end{equation}
with a potential 
\begin{equation} \label{eq:Pot}
V(r)= f\left[ \frac{(d-1)}{4 r} \frac{(f h)'}{f} + \frac{(d-1)(d-3)}{4} \frac{h}{r^2} + \frac{l(l+d-2)}{r^2} +m^2 \right] .
\end{equation}
We examine the form of the potential in global AdS and AdS-Schwarzschild:
\begin{figure}[tbp] 
\centering
\subfigure[]{
	\includegraphics[width=2.7in]{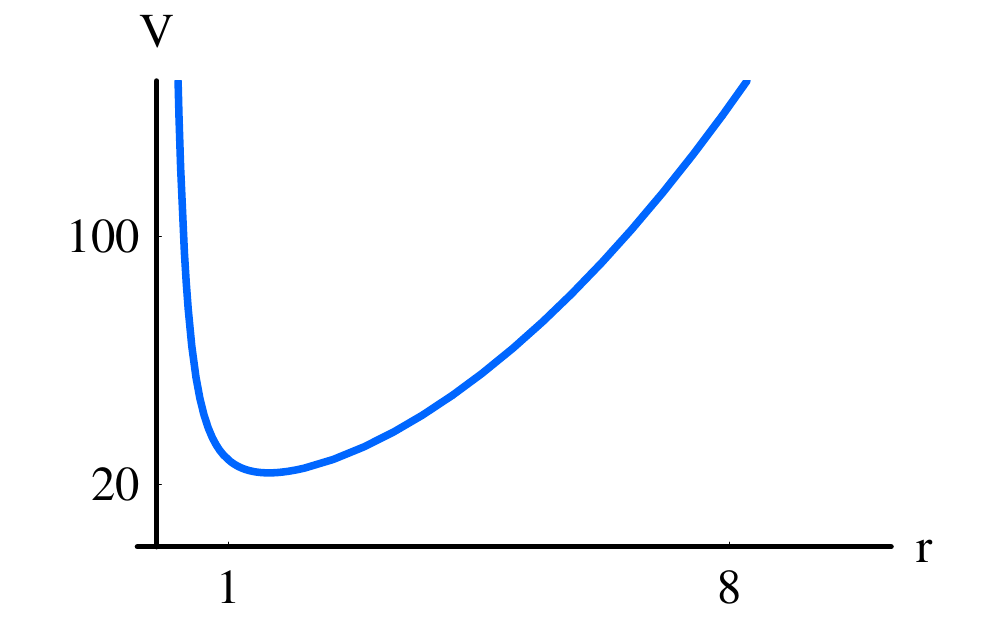}
	}
	\hspace{.2in}
		\subfigure[]{
	\includegraphics[width=2.7in]{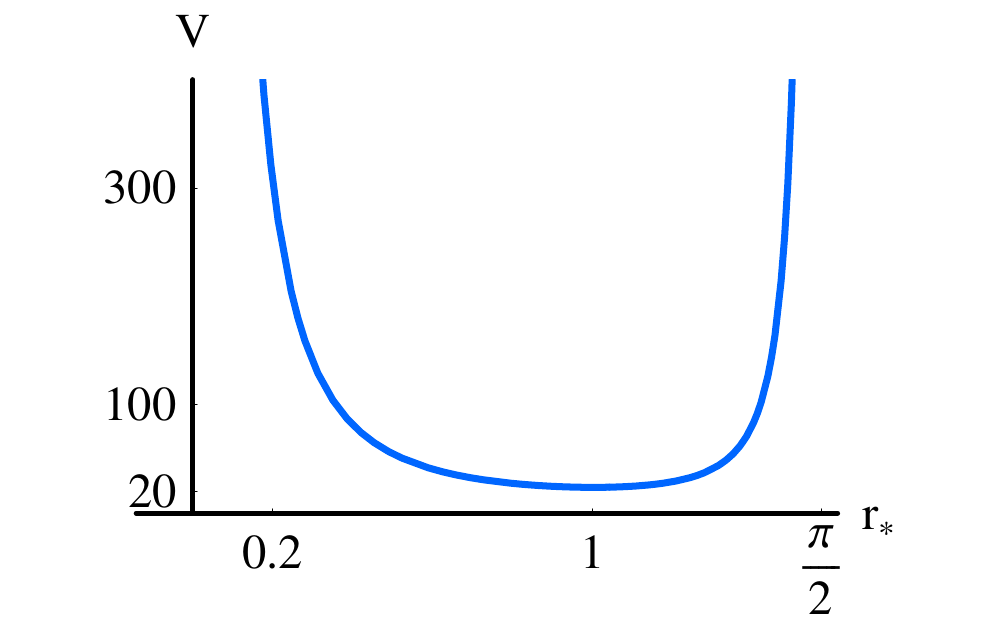}}
\caption{The wave equation can be recast as a Schr\"odinger equation (\ref{eq:Schro}). We plot the global AdS$_4$ potential (\ref{eq:VGlob}) for $l=3$ for a massless field. The plot on the left is in terms of the radial coordinate $r$ appearing in the AdS metric (\ref{eq:Global}). The plot on the right is  in terms of the tortoise coordinate $r_*$, and is the one relevant for solving (\ref{eq:Schro}). The two are related through $r = \tan r_*$. The tortoise coordinate has the effect of compressing the potential at large $r$, while leaving small $r$ unaffected. The AdS barrier occurs at $r_*$ very close to $\pi/2$; its narrowness allows the modes to decay only as a power law: $\phi \sim r^{-\Delta}$. } \label{fig:Global}
\end{figure}

\subsubsection*{Global AdS}

Global $AdS_{d+1}$ has the metric
\begin{equation} \label{eq:Global}
ds^2 = - (1+r^2)dt^2 + \frac{dr^2}{1+r^2} +r^2 d\Omega_{d-1}^2 ,
\end{equation}
and correspondingly a potential 
\begin{equation} \label{eq:VGlob}
V_{Global}(r) = (1+r^2)\left[ \frac{d^2 -1}{4} + m^2 + \frac{(d-1)(d-3)+4 l(l+d-2)}{4 r^2}\right] .
\end{equation}

The potential for global AdS is plotted in Fig.~\ref{fig:Global}. The  potential is dominated at small $r$ by the angular momentum barrier $l(l+d-2)/r^2$, and at large $r$ by the AdS barrier proportional to $r^2$. At intermediate radius, these terms balance and the potential attains a minimum set by the angular momentum $l$. The minimum of the potential, which at large $l$ is approximately $l(l+d-2)$, sets the lower bound on $\omega$.

\subsubsection*{AdS-Schwarzschild}

\begin{figure}[tbp] 
\centering
\subfigure[]{
	\includegraphics[width=2.7in]{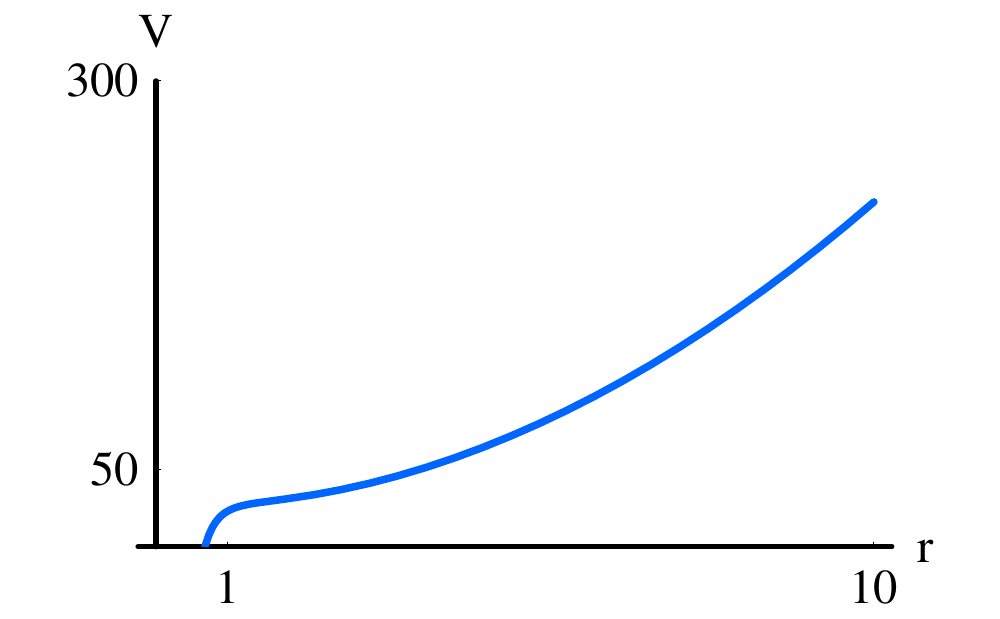}
	}
	\hspace{.2in}
		\subfigure[]{
	\includegraphics[width=2.7in]{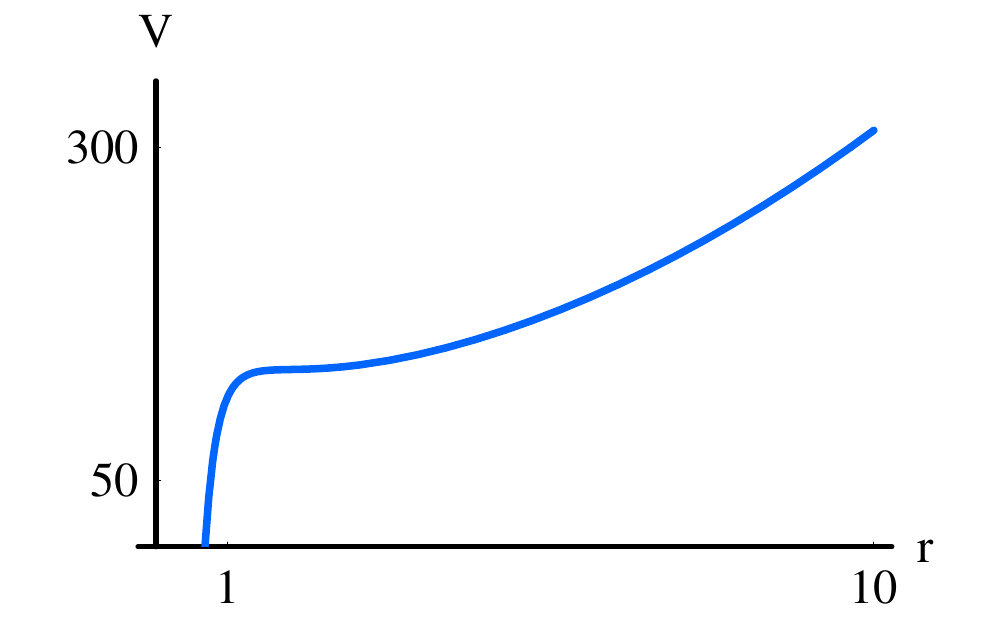}}
\caption{A plot of the $AdS_4$-Schwarzschild ($r_0=1$) potential (\ref{eq:VBH}) as a function of the radial coordinate $r$. The plot on the left is for $l=4$, and the one on the right for $l = 10$. Unlike for pure AdS, $\omega$ is not bounded from below by $l$; for a fixed $\omega$, $l$ can be arbitrarily high. The barrier an $\omega$ mode must pass through grows as $l$ increases. This results in the $\omega \ll l$ modes having exponential behavior in $l$. Intuitively, these modes become ever more confined near the horizon with increasing $l$. } \label{fig:BH}
\end{figure}

AdS- Schwarzschild is of the form
\begin{equation} \label{eq:Sch}
ds^2 = - \left(1+ r^2 - \left(\frac{r_0}{r}\right)^{d-2}\right)dt^2 + \frac{dr^2}{1+ r^2 - \left(\frac{r_0}{r}\right)^{d-2}} + r^2 d\Omega_{d-1}^2 .
\end{equation}
giving a potential 
\begin{equation} \label{eq:VBH}
V_{BH}(r) = \left[1+ r^2 - \left(\frac{r_0}{r}\right)^{d-2}\right]\left[ \frac{d^2 -1}{4} + m^2 +\frac{(d-1)^2}{4} \frac{r_0^{d-2}}{r^{d}} + \frac{(d-1)(d-3)+4 l(l+d-2)}{4 r^2} \right] .
\end{equation}

The potential for  AdS-Schwarzschild is shown in Fig.~\ref{fig:BH}. For large $r$ the behavior is the same as for pure AdS. However, the behavior is different for $r$ close to the horizon: the factor of $f$ vanishes at the horizon, forcing the potential to vanish as well. The vanishing of the AdS-Schwarzschild potential at the horizon allows $\omega$ to be arbitrarily small, regardless of the value of $l$. This is in contrast with pure AdS. 

\subsubsection*{BTZ}
A nonrotating BTZ black hole has a metric similar to AdS-Rindler (\ref{eq:AdSRindler}). In the form of (\ref{eq:Static}), $f(r) = h(r) = r^2-M$, giving a potential
\begin{equation}
V_{BTZ}(r) = (r^2-M) \left(\frac{3}{4} + m^2 + \frac{l^2 +M/4}{r^2}\right).
\end{equation}
The BTZ potential has similar properties to that of AdS-Schwarzschild.

\section{Black hole smearing functions and large angular momentum modes} \label{sec:4}

In this section we explain why global AdS admits a smearing function while AdS-black hole backgrounds do not. The reason is simple: if a black hole is present, modes with $l \gg \omega$ (and $l$ arbitrarily large) become allowed. These modes are highly suppressed at large $r$ by the centrifugal barrier. An attempt to calculate the smearing function via a mode sum immediately gives a divergence when performing the sum over $l$ at a fixed $\omega$. We will show there is no smearing function in two ways. First in Sec.~\ref{sec:4.1} we show the existence of these $l \gg \omega$ modes in itself, independent of the details of the metric, prevents a smearing function for bulk points at large $r$. In Sec.~\ref{sec:4.2} we use WKB to directly solve for the modes, showing there is no smearing function for any bulk point $B$. We should note that there are other cases where a smearing function fails to exist, even without a horizon and the associated $l \gg \omega$ modes, as we will show in Sec.~\ref{sec:5}.

In Sec.~\ref{sec:4.1} we review how in pure AdS modes oscillate as $e^{i q z}$ where $z=0$ is the boundary and $q^2 = \omega^2 - l(l+d-2)$. If modes existed with $q^2<0$, then they would grow exponentially  as $e^{\kappa z}$ where $\kappa^2 = - q^2$. Since black hole backgrounds asymptotically approach pure AdS, their $q^2 <0$ modes will display this exponential behavior in $l$. Connecting with the discussion in Sec.~\ref{sec:2}, this means the sum in (\ref{eq:K}) will not converge, and hence these modes forbid a smearing function for bulk points near the boundary. 

In Sec.~\ref{sec:4.2} we show that in the limit of  high $l$, the potential (\ref{eq:Pot}) considerably simplifies, with only the centrifugal barrier remaining. Using WKB we solve to find the modes. The result shows an exponential behavior in $l$ for these $l\gg \omega$ modes, for any bulk point. Thus we find there is no smearing function for any point in any static spherically symmetric spacetime with a horizon. 

\subsection{Asymptotic behavior of the wave equation} \label{sec:4.1}
For an asymptotically AdS space, at large $r$ the $f(r),\ h(r)$ in (\ref{eq:Static}) have the limit $f(r),\ h(r) \rightarrow r^2$. Changing variables to $z = 1/r$, we write the metric as
\begin{equation} \label{eq:Poin}
ds^2 = \frac{-dt^2 +dz^2 + d \Omega^2}{z^2}.
\end{equation}
For small angles  (\ref{eq:Poin}) resembles the metric of the Poincare Patch. The wave equation for the radial modes is
\begin{equation} \label{eq:phiz}
z^2 \phi'' - z(d-1) \phi' + (z^2 q^2 -m^2)\phi = 0,
\end{equation}
where $q^2 = \omega^2 - l(l+d-2)$. Substituting $\phi(z) = z^{d/2}\psi(z)$ yields
\begin{equation} \label{eq:Poincare}
z^2 \psi '' + z \psi' +( z^2 q^2 -\nu^2) \psi =0 ,
\end{equation}
where we defined $\nu^2 = m^2+ d^2/4$. For $\omega^2>l (l+d-2)$ this gives $\psi(z) = J_{\nu}(q z)$ and hence
\begin{equation} \label{eq:PoinSol}
\phi(z) = z^{d/2} J_{\nu}(q z) ,
\end{equation}
which resembles the usual solution in the Poincare Patch.\footnote{The other solution, $Y_{\nu}(qz)$, is discarded because it doesn't have the correct behavior $\phi \rightarrow z^{\Delta}$ near the boundary that is expected of a normalizable mode.} Modes with $\omega^2< l(l+d-2)$ have negative $q^2$. Of course these modes don't exist in pure AdS, but they do in AdS-Schwarzschild. Defining $\kappa^2 = -q^2$, we get $\psi(z) = J_{\nu}(i \kappa z) \equiv e^{i \nu \pi/2} I_{\nu}(\kappa z)$.\footnote{In the context of \textit{Euclidean} AdS/CFT in the Poincare Patch one has this scenario of $q^2<0$. There are two solutions: $I_{\nu}(\kappa z)$ and $K_{\nu}(\kappa z)$. The $K_{\nu}(\kappa z)$ solution is kept while the $I_{\nu}(\kappa z)$ is discarded precisely because of its exponential growth in the bulk. Of course, for us $K_{\nu}(\kappa z)$ cannot be kept since it grows exponentially as the boundary is approached and so is nonnormalizable.} Since the Bessel function $J_{\nu}(x)$ oscillates, $I_{\nu}(x)$ grows exponentially. We can see the exponential growth directly. Letting $\phi(z) = z^{\frac{d-1}{2}} u(z)$, (\ref{eq:phiz}) becomes

\begin{equation} \label{eq:uSchr}
u'' - \left[\kappa^2 +\frac{\nu^2 - 1/4}{z^2}\right]u =0~.
\end{equation}
In the limit of $z \gg \sqrt{\nu^2 - 1/4}/\kappa$,\ Eq.~\ref{eq:uSchr} is solved by $u = e^{\kappa z}$. Thus, we see that for large $l$ and $l \gg \omega$ the modes behave as
\begin{equation} \label{eq:bigl}
\phi(z) = z^{\frac{d-1}{2}} e^{lz}.
\end{equation}

Our use of the pure AdS metric is justified for sufficiently small $z$. However for any $z>0$, there exists an $l$ sufficiently large such that $z\gg \sqrt{\nu^2 - 1/4}/\kappa$ \ is satisfied. Since computing a smearing function involves summing over $l$ arbitrarily large, we are guaranteed to reach regime (\ref{eq:bigl}) at sufficiently high $l$. 

\subsubsection*{Smearing function for static spherically symmetric spacetimes}
In the case of static spherically symmetric spacetimes, the solutions (\ref{eq:phi}) can be inserted into the smearing function (\ref{eq:K}), giving

\begin{equation} \label{eq:Ksphere}
K(r,t, \Omega| t',\Omega') = \int{d\omega e^{i \omega (t-t')} \sum_{l, m_i}{\phi_{\omega, l}(r) Y_{l m_i}(\Omega)Y_{l m_i}^{*}(\Omega ')}}~,
\end{equation}
where $m_i$ denotes all the angular quantum number, $m_1,...,m_{d-2}$, and we have normalized the time dependent piece with respect to the boundary Klein-Gordon norm. The radial modes $\phi_{\omega, l}$ are solutions to the radial wave equation (\ref{eq:Schro}) and should be normalized so that $\phi_{\omega l} \rightarrow r^{- \Delta}$ as $r\rightarrow \infty$. For AdS-Schwarzschild, the energies $\omega$ are continuous and so we have written an integral over $\omega$; for global AdS this would be replaced by a discrete sum over $n$ as $\omega_n = 2n+ l + \Delta$.

If only modes with $\omega> \sqrt{l (l+d-2)}$ are allowed then, as we saw above, the near boundary solution (\ref{eq:PoinSol}) is, when properly normalized, 
\begin{equation}
\phi_{\omega l} (r)  = 2^{\nu} \Gamma(\nu +1) \frac{ J_{\nu}(q/r)}{r^{d/2}q^{\nu}} .
\end{equation}
Inserting the $\phi_{\omega l}(r)$  into (\ref{eq:Ksphere}), we see the sum converges. On the other hand, for modes with $\omega< \sqrt{l (l+d-2)}$, and in particular the high $l$ ones with solution (\ref{eq:bigl}), the sum over $l$ in (\ref{eq:Ksphere}) is hopelessly divergent.

\subsection{Large angular momentum and WKB} \label{sec:4.2}
Our goal here is to directly show the exponential behavior in $l$ of the modes $\phi_{\omega l}(r)$ for large $l$ and $l\gg\omega$ for any bulk point. The smearing function doesn't exist due to the modes with arbitrarily large $l$, which is why this is a sufficient limit to consider. We will also see how the details of the metric become irrelevant in the large $l$ limit, with the centrifugal barrier dominating the potential (\ref{eq:Pot}).

Modes with energy $\omega$ have a turning point at $r=r_t$ which satisfies $V(r_t) =\omega^2$. In the limit of $l\gg \omega$, the turning point approaches the horizon, $r_t \approx r_h$. For $r>r_t$ the modes always have $\omega^2 < V$ and thus decay. For $r>r_t$, we can use  WKB:

\begin{equation} \label{eq:WKB}
u(r) = \frac{1}{\sqrt{p}} \exp{\left(- \int_{r_t}^{r}{p ~dr_*}\right)},
\end{equation}
where $p^2 = V- \omega^2$.

We will only be interested in the exponential term, so we drop the $1/\sqrt{p}$ prefactor. As discussed in Sec.~\ref{sec:2}, in order to compute the smearing function we need to normalize all the bulk modes so that their boundary limit (upon stripping off $r^{\Delta}$) is normalized with respect to the boundary norm. In terms of $u(r)$, we need its coefficient to approach $1$ as $r\rightarrow \infty$. 
Thus, 
\begin{equation} \label{eq:u1}
u(r) = \exp\left(\int_{r}^{\infty}{\frac{dr'}{\sqrt{f(r') h(r')}}}\sqrt{V(r') - \omega^2} \right) ,
\end{equation}
where we used the relation between the radial coordinate and the tortoise coordinate, $dr_* = dr/\sqrt{f h}$. 

The key point is that for any point outside the horizon, $r>r_h$, there is an $l$ sufficiently large such that the potential (\ref{eq:Pot}) can be approximated by 
\begin{equation} \label{eq:approxPot}
V(r) = f \frac{l^2}{r^2},
\end{equation}
where for simplicity we used $l(l+d-2) \approx l^2$. In (\ref{eq:u1}) it is sufficient to only integrate for some finite distance $\delta$ away from $r$ to see the exponential behavior in $l$,
\begin{equation}
u(r) > \exp{\left(\int_r^{r+\delta}{\frac{dr'}{\sqrt{f(r')h(r')}} \sqrt{V(r') - \omega^2}} \right)} .
\end{equation}
For any $\delta$ we want, there is an $l$ sufficiently large such that the potential (\ref{eq:Pot}) can be approximated by (\ref{eq:approxPot}) for all radii between $r$ and $r+\delta$. Thus, using the approximate potential (\ref{eq:approxPot}) and neglecting $\omega^2$ we get,
\begin{equation} \label{eq:expGrowth}
u(r) > \exp{\left(l \int_{r}^{r+\delta}\frac{dr'}{r'\sqrt{h(r')}}\right)}.
\end{equation}
This demonstrates the exponential growth in $l$ of the modes. This is true for any bulk point $r$; the only difference is the larger $r$, the greater the $l$ before the exponential growth (\ref{eq:expGrowth}) sets in.

In the limit of large $r$ we can approximate $f(r) \approx h(r) \approx r^2$. This yields $u(r) \rightarrow e^{l/r}$. Recalling $\phi = u/ r^{(d-1)/2}$, this reproduces (\ref{eq:bigl}). Additionally, (\ref{eq:expGrowth}) matches the exponential growth in $l$ of the exact hypergeometric function solution (\ref{eq:hyp}) found in \cite{Kesk} for the BTZ black hole. 

\section{Smearing functions for other spacetimes} \label{sec:5}

We have seen if there is a horizon the potential (\ref{eq:Pot}) vanishes at the horizon and consequently any frequency $\omega>0$ is allowed. The arguments of the previous section show there is no smearing function. In this section we examine more generally when a smearing function exists. In Sec.~\ref{sec:5.1} we consider a general static spherically symmetric spacetime (\ref{eq:Static}), and find a simple criteria on the metric (\ref{eq:derivCond}) which gaurantees there will be modes with exponential behavior in $l$ for high $l$, and hence there will not be a smearing function for some bulk points. In Sec.~\ref{sec:5.2} we  examine the possibility of trapped null geodesics as a proxy for a smearing function not existing for an arbitrary spacetime. In the special case of static spherically symmetric spacetimes we demonstrate that the existance of trapped null geodesics prevents a smearing function.

\subsection{Static spherically symmetric spacetimes} \label{sec:5.1}
In this section we examine the existence of a smearing function for spacetimes of the form (\ref{eq:Static}) which do not possess horizons. 

\begin{figure}[tbp] 
\centering
\subfigure[]{
	\includegraphics[width=2in]{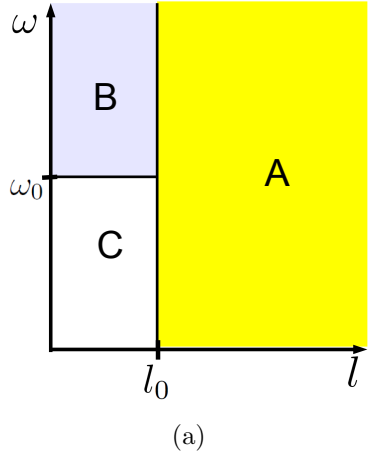}
	}
\caption{ We are interested in finding for which static spherically spacetimes without horizons a smearing function exists. The smearing function involves the sum (\ref{eq:Ksphere}) over modes, which can be grouped into $3$ different regimes. Only $A$ posses a threat to the convergence of (\ref{eq:Ksphere}). At large $r$ the metric, and consequently the potential (\ref{eq:Pot}), looks like that of pure AdS (\ref{eq:Vbigr}). At smaller $r$, in regime $A$ the angular momentum $l$ is so large that all terms in the potential except for the centrifugal barrier (\ref{eq:Vbigl}) are irrelevant.} \label{fig:Regions}
\end{figure}

The question of the existance of a smearing function is the question of the convergence of the sum (\ref{eq:Ksphere}) over $\omega$ and $l$ at a given value of $r$. To answer this question, we will need to estimate the size of each mode with a given $\omega$ and $l$, for every $\omega$ that is allowed. It is convenient to divide the $(\omega,l)$-plane into three regions, shown in Fig.~\ref{fig:Regions}, according to the sizes of $\omega$ and $l$ relative to certain large values $\omega_0$ and $l_0$ which depend only on $f$ and $h$ and will be defined carefully below. Region $A$ consists of all modes with $l>l_0$, region $B$ consists of all modes with $\omega> \omega_0$ and $l<l_0$, and region $C$ consists of the remaining modes with $\omega <\omega_0$ and $l<l_0$. 

\subsubsection*{Approximating the potential}
\begin{figure}[tbp] 
\centering
\subfigure[]{
	\includegraphics[width=4.5in]{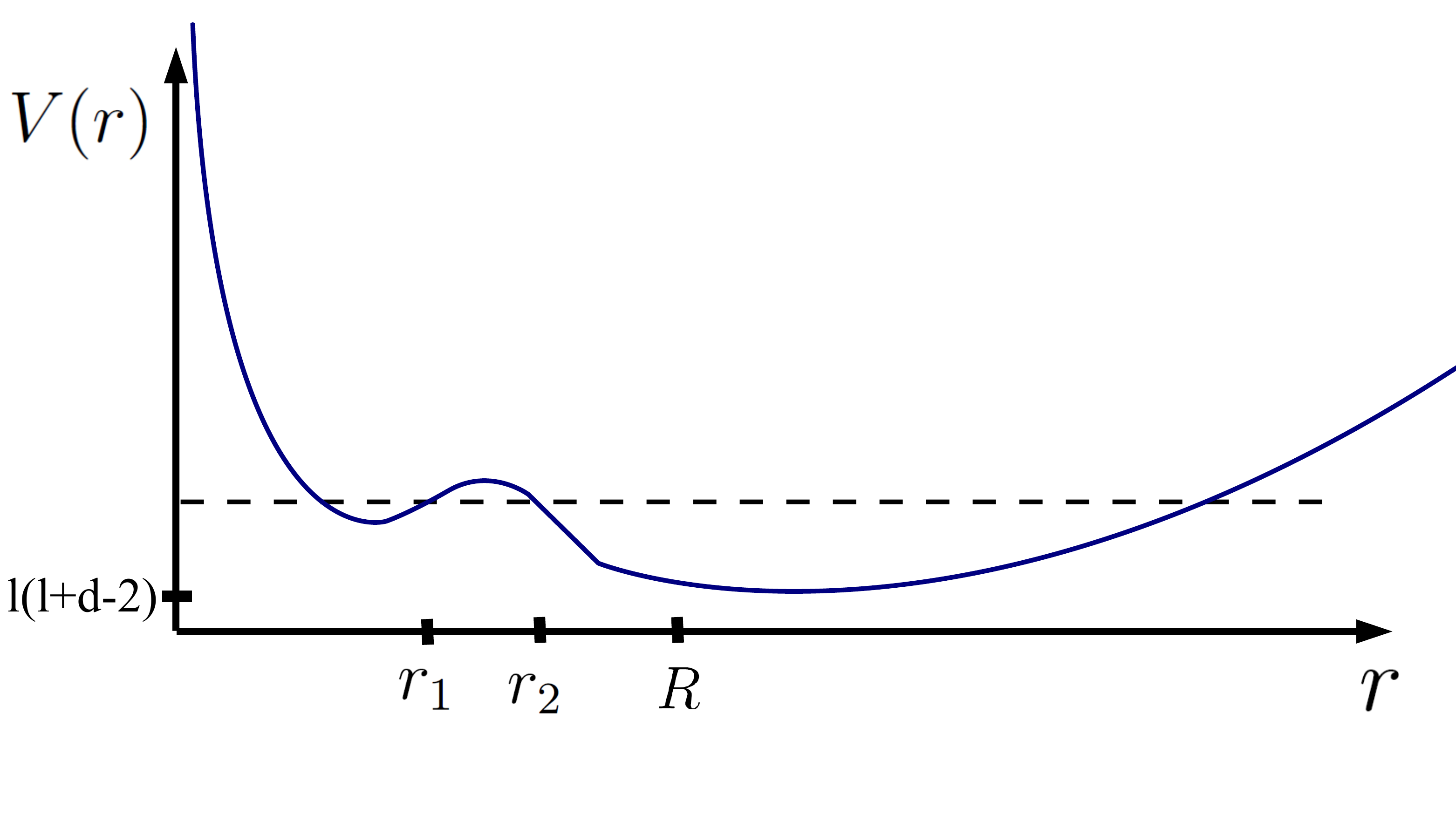}
	}
\caption{ The wave equation can be recast as a Schr\"odinger equation (\ref{eq:Schro}) with a potential $V(r)$ and an energy $\omega^2$. Here we sketch a possible potential (\ref{eq:Pot}) for which a smearing function doesn't exist. At large $r$, $r>R$, the the potential looks like that of pure AdS (the figure has been compressed; the distance between $r_2$ and $R$ is really much larger). At smaller $r$ the potential, for large $l$, is approximated by (\ref{eq:Vbigl}). If $f(r)/r^2$ ever has positive slope, as shown above, some of the modes $\omega$ (dashed line) will have to tunnel through the barrier. Consequently, the sum (\ref{eq:Ksphere}) will diverge for  $r<r_2$.}\label{fig:PotSketch}
\end{figure}

To aid our calculation, we will approximate the behavior of the potential for large, small, and intermediate values of $r$. 

\textbf{Large $r$}

Since the metric approaches pure AdS at large $r$, we can approximate $f(r) \approx h(r) \approx r^2$ for $r>R$, where $R$ is some sufficiently large radius that depends on $f$ and $h$. The potential for $r>R$ thus takes the form
\begin{equation} \label{eq:Vbigr}
V(r) \approx l (l+d-2) + \left(\frac{d^2-1}{4} + m^2\right) r^2,  \ \ \  r>R . 
\end{equation}

\textbf{Small $r$}

As $r\to 0$, $h(r) \to 1$ to avoid a conical singularity, and $f(r) \to f_0>0$. Thus we can find some small $\epsilon>0$ such that 
\begin{equation}
V(r) \approx f_0\left(\frac{l(l+d-2) + (d-1)(d-3)/4}{r^2}\right), \  \ \ \  \ \ \ r<\epsilon.
\end{equation}
Note that the form of the potential implies that $\omega$ is quantized. Aside from the constant $f_0$, the potential for $r<\epsilon$ does not depend on the details of the geometry.

\textbf{Intermediate $r$}

For $\epsilon<r<R$ it will be useful to do a separate analysis for modes lying in the three different $(\omega,l)$-regions shown in Fig.~\ref{fig:Regions}.

\textbf{A: }Since $f$, $h$, and their derivatives are bounded functions for $r<R$, we can find an $l_0$ sufficiently large so that for all $l>l_0$ all terms in the potential (\ref{eq:Pot}) except the centrifugal barrier are irrelevant for all $r<R$,

\begin{equation} \label{eq:Vbigl}
V(r) \approx f(r) \frac{l(l+d-2)}{r^2} \ , \ \ \ \ \ \  r<R \  \ \ \ \text{and}  \ \ \ \ \ l>l_0.
\end{equation}
This potential agrees with our small-$r$ approximation above when $r<\epsilon$. Note that the potential in this region has an overall scaling with $l$.

\textbf{B:  } We choose an $\omega_0$ sufficiently large such that $\omega_0^2 \gg V(r)$ for all $\epsilon <r < R$ and $l<l_0$. For modes in region $B$ the potential is negligible at intermediate values of $r$.

\textbf{C:} For modes in region $C$ all of the details of the potential are important, and there is no useful approximation.

\subsubsection*{Convergence of sum for smearing function}

Since $\omega$ and $l$ are quantized, there are only a finite number of modes in region $C$, so that part of the sum (\ref{eq:Ksphere}) converges. In region $B$ the modes experience the same potential as in pure AdS (aside from an inconsequential scaling of $f_0$ at small $r$), so that part of the sum will converge as well. This only leaves region $A$ to analyze. 

Let us suppose the potential (\ref{eq:Vbigl}) has positive slope for some range of $r$, 
\begin{equation} \label{eq:derivCond}
\frac{d}{dr} \left(\frac{f(r)}{r^2}\right) > 0 \ \ \ \ \ \ \text{for some} \ \  r.
\end{equation}
We will now show if this occurs then there is no smearing function for some bulk points due to an exponential divergence in region $A$.  In Fig.~\ref{fig:PotSketch} we sketch an example of potential for which (\ref{eq:derivCond}) occurs. Consider the limit of large $\omega$ and large $l$. This is the classical limit of the Schr\"odinger equation (\ref{eq:Schro}), as can be seen from the fact that the range of $r_*$ is finite and fixed, while the potential $V(r)$ and energy $\omega$ are getting large. Thus, it is guaranteed that there exists a mode $\omega$ lying within any classically allowed energy interval. If (\ref{eq:derivCond}) is satisfied in a neighborhood of $r=r_1$, then that neighborhood consists of classical turning points for an interval of possible values of $\omega$.  Let $r_2 >r_1$ be any point in the classically forbidden region for these values of $\omega$.\footnote{We treat $r_1$ and $r_2$ as if they are less than $R$ for the puprose of approximating the potential. However, if they are larger than $R$ it makes little difference. In the large-$l$ limit the extra terms in (\ref{eq:Vbigr}) become irrelevant for $r<r_2$.} Then the field at any $r$ in the range $r_1< r < r_2$ (or any $r$ in the classically allowed region $r<r_1$)  is larger than that at $r_2$ by a WKB factor of 
\begin{equation} \label{eq:WKBfac}
\exp\left({\int_{r}^{r_2}\frac{dr}{\sqrt{f h}}{\sqrt{V- \omega^2} }}\right) = \exp\left(l \int_{r}^{r_2}{\frac{dr}{\sqrt{fh}} \sqrt{\frac{f}{r^2} - \left(\frac{\omega}{l}\right)^2}}\right).
\end{equation}

There is a subtlety here: in addition to this decaying solution, there is also an exponentially growing solution and the eigenstate will in general be a linear combination of the two if there is a second classically allowed region when $r>r_2$ (as in the scenario of Fig.~\ref{fig:PotSketch}). If both solutions contribute with comparable coefficients, then the eigenstate will not be exponentially larger at $r$ than it is at $r_2$ as we have claimed. That phenomenon occurs, for instance, in the symmetric double well potential familiar from quantum mechanics. However, that kind of behavior is special to the symmetric, degenerate case. As long as the energy differences between approximate eigenstates localized on either side of the barrier is larger than the exponentially small tunneling factor, the true eigenstates of the system will be exponentially well localized on one side of the barrier, and we can restrict attention to those localized in the $r<r_1$ region.

It is clear that (\ref{eq:WKBfac}) can be made arbitrarily large by making $l$ large.  Speficially, let $\alpha = \omega/l$ where $\omega$ and $l$ are the modes considered above which are suppressed and give behavior (\ref{eq:WKBfac}). Now consider the portion of the sum (\ref{eq:Ksphere}) concentrated on the line of fixed $\alpha$. Thus, (\ref{eq:Ksphere}) will not converge and there will not be a smearing function for points $r<r_2$. We should note that unlike the black hole case considered earlier, which did not have a smearing function for any bulk point, for a potential like in Fig.~\ref{fig:PotSketch} there is a smearing function for bulk points at large $r$.

A remaining question is the converse of our statement: if $f(r)/r^2$ has non-positive slope for all $r$, is the smearing function guaranteed to exist? In this case there are no turning points at intermediate values of $r$, and hence no opportunity for exponential WKB factors. However, it is possible that the magnitude of the slope of $f(r)/r^2$ is small for some range of $r$, and then the $1/\sqrt{p}$ factor in (\ref{eq:WKB}) can become large. It is conceivable that one could still make sense of (\ref{eq:Ksphere}), despite power law growth in the summand, through regulation and analytic continuation. This requires further analysis, and is something we intend to investigate in subsequent work.

\subsection{Trapped null geodesics} \label{sec:5.2}
We have established a smearing function does not exist in a black hole background. In a general static spherically symmetric spacetime, we have shown that it does not exist if the metric has the property (\ref{eq:derivCond}). In a more general spacetime without a high degree of symmetry, the mode sum approach to constructing a smearing function is inapplicable. This motivates us to search for a simply proxy for the existence of a smearing function. In this section we explore the following proposal: there is a smearing function iff all null geodesics have at least one endpoint on the boundary. 

A smearing function allows one to make the statement (\ref{eq:Kabat}) about the mapping between bulk and boundary operators. However, finding a smearing function is a classical field theory problem. At the level of individual modes we saw a smearing function for a point $B$ in the bulk fails if there are modes whose imprint on the regulated boundary is exponentially small compared to their value $B$. Throughout this paper we held the boundary imprint fixed and saw the value at $B$ grow arbitrarily large, causing (\ref{eq:K}) to diverge. Keeping the field value at $B$ fixed, this would correspond to modes with boundary imprint becoming arbitrarily small.

Since field configurations are built out of modes, we can state this as: a smearing function fails to exist if there are bulk field solutions $\Phi(B)$ with arbitrarily small boundary imprint. When one takes the geometric optics limit, field solutions become arbitrarily well localized along null geodesics. At a heuristic level, this motivates the simple criterion of trapped null geodesics, which we will now explore quantitatively..\footnote{In ~\cite{BarLeb92,Tat04} it was shown for a large class of hyperbolic differential equations that the diagnostic if reconstruction of a bulk field depends on boundary data continuously (for a particular choice of boundary norm) is that there not be any trapped null geodesics. In \cite{BouFre12} null geodesics were explored in the context of subregion dualities due to these results and with the motivation of establishing if a collection of boundary observes can physically reconstruct the bulk field in a subregion of AdS. This question is of secondary concern to us here; our interest is rather in the nature of the bulk-boundary dictionary (if (\ref{eq:Kabat}) is possible). }

\subsubsection*{Geodesic equation}

To find the motion of null geodesics in the spacetime (\ref{eq:Static}), we note  that the timelike Killing vector gives the conserved quantity $E = f \dot{t}$, and the Killing vector in one of the angular directions, $\theta$, gives $L= r^2 \dot{\theta}$. Here we are using the notation $\dot{x}^{\mu} \equiv dx^{\mu}/d\lambda$ where $\lambda$ is the affine parameter.

\begin{figure}[tbp] 
\centering
\subfigure[]{
	\includegraphics[width=2.7in]{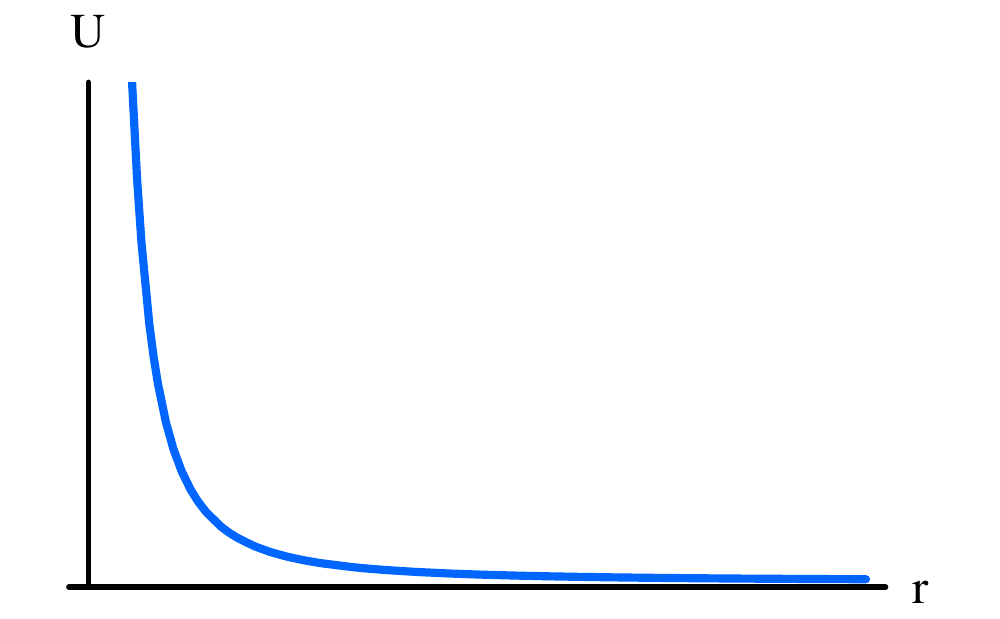}
	}
	\hspace{.2in}
		\subfigure[]{
	\includegraphics[width=2.7in]{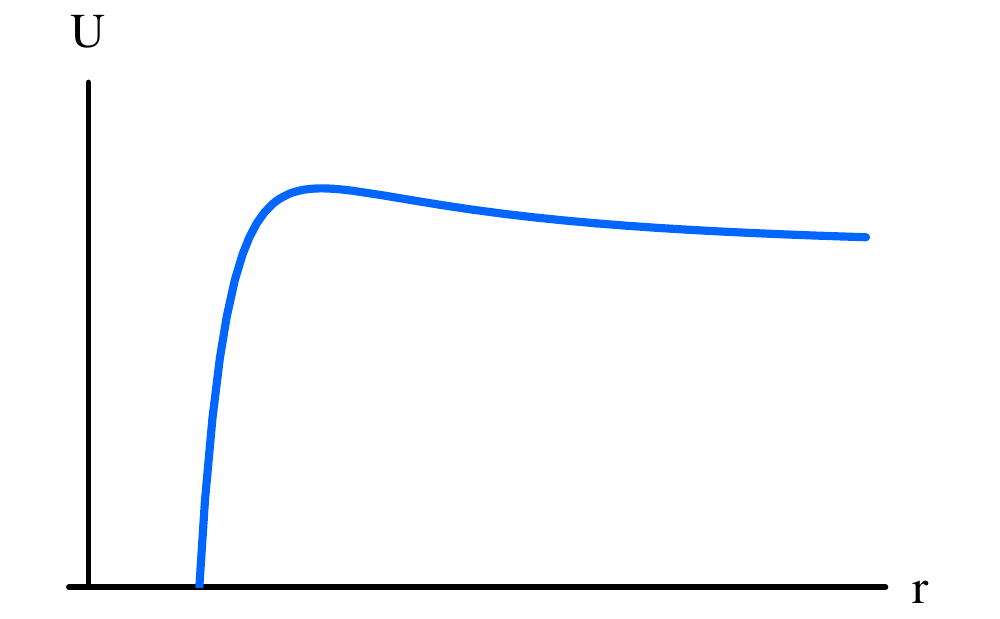}}
\caption{The equation for a null geodesic is that of a particle traveling in a 1-d potential (\ref{eq:U}). The potential is plotted for (a) pure AdS and (b) AdS-Schwarzschild ($M=1$ in AdS$_4$). In pure AdS all null geodesics have an endpoint on the boundary, as can be seen from the figure on the left. This is in contrast to spacetime with horizons (right figure) which have some null geodesics which are trapped as a result of the potential $U$ vanishing at the horizon. More generally, whenever there are trapped null geodesics, then there is no smearing function for some points in the bulk.} \label{fig:NullGeodesics}
\end{figure}

The geodesic equation can be written as 
\begin{equation} \label{eq:Geodesic}
\frac{f}{h} \dot{r}^2 + L^2 \frac{f}{r^2} = E^2 .
\end{equation}
This is just the Newtonian energy conservation equation for a particle with position-dependent mass moving in a potential
\begin{equation} \label{eq:U}
U = L^2 \frac{f}{r^2} \ .
\end{equation}
In black hole backgrounds the potential (\ref{eq:U}) vanishes at the horizon, leading to trapped null geodesics. 
More generally, (\ref{eq:U}) tells us there are no trapped geodesics iff $U'<0$ for all $r$.

It is interesting to note that even a small dense star in AdS can have trapped null geodesics. All the star needs is to have a radius $\mathcal{R}$ which lies in the range $2M < \mathcal{R}<3M$. The metric for $r>\mathcal{R} $ is described by the Schwarzschild metric. Since for the Schwarzschild metric $U'(r)>0$ for $r<3M$, null geodesics will get trapped at small $r$.  

\subsubsection*{Trapped null geodesics $\Rightarrow$  no smearing function}

In 5.1 we saw that for the question of the existence of a smearing function, only the behavior of the high $l$ modes was relevant. In this regime (labeled $A$ in Fig.~\ref{fig:Regions}) the potential in the Schr\"odinger equation for the modes was well approximated by (\ref{eq:Vbigl}). Yet this is exactly the same as the classical particle potential (\ref{eq:U}) for a null geodesic. Our condition for a smearing function not existing (\ref{eq:derivCond}) is the same as the condition for the existence of trapped null geodesics. Thus we conclude that if there are trapped null geodesics, then a smearing function does not exist.

We note that since null geodesics are only sensitive to the local metric, trapped null geodesics cannot tell us for which regions of the bulk there is no smearing function. If a null geodesic is confined to $r<r_t$, this indicates there is no smearing function for $r<r_t$, but it says nothing about a smearing function for $r>r_t$. The existence of a smearing function for $r > r_t$ depends, as we explained above, on the existence of additional classically allowed regions with $V(r) < \omega^2$ for $r>r_t$. A classical null geodesic confined to $r<r_t$ cannot probe these aspects of the potential.

\section{Conclusions}
In this paper we have further explored one of the approaches to establishing the dictionary between bulk and boundary operators. In this approach, a bulk operator is expressed in terms of bulk operators at asymptotically large radius, which are then mapped to local boundary operators through (\ref{eq:DIC}). While this approach works in pure global AdS, we have argued it can fail if there are bulk modes which have an arbitrarily small tail at large radius. We have shown that AdS-Schwarzschild backgrounds are a case where this smearing function approach fails as a result of modes with arbitarily high angular momentum $l$, but fixed boundary energy, $\omega$. 

Understanding in general when a spacetime has a region for which a bulk operator cannot be expressed in terms of smeared local boundary operators remains an important future problem.  We have shown that for static, spherically symmetric spacetimes, this question can be answered by considering the behavior of modes with large $\omega$ and large $l$. These modes satisfy a Schr\"odinger-like equation with a potential that is the same as the potential experienced by classical null geodesics. These results suggest that a smearing function may not exist for some bulk points in any spacetime which has trapped null geodesics .

The extent to which the absence of a smearing function modifies the bulk-boundary dictionary remains to be seen. It is possible one can obtain an approximate smearing function by imposing a cutoff in the bulk and excluding high $l$ and high $\omega$ modes from the sum (\ref{eq:Ksphere}) defining the smearing function. Additionally, as we discussed, the existance of a smearing function is a more stringent requirement than simply being able to reconstruct a bulk field solution given some particular boundary data. To this extent, even though (\ref{eq:Smearing}) may not exist, (\ref{eq:Phi2}) exists at $N=\infty$. However, unlike a smearing function, it is unclear that (\ref{eq:Phi2}) can be generalized to situations with broken spherical and time-translation symmetry, and so may be of limited use. Another option would be to try to construct a smearing function which uses the complexified boundary, as done in ~\cite{HamKab06} for AdS-Rindler, and perhaps this would gives clues as to what the real spacetime representation of the bulk-boundary map is. 

When a smearing function does exist, it means that bulk data provided at large radius for a sufficient time extent completely determines the bulk everywhere. In a way, it makes holography seem less powerful; a spatial direction has just been replaced by a time direction. Of course, the power of AdS/CFT is due to the CFT Hamiltonian which one can use to evolve the right side of (\ref{eq:K}) to a single time. The resulting operator is highly nonlocal and known as a precursor ~\cite{PolSus99}: an operator which encodes what happened deep in the bulk long before casuality allows a local operator $O(b)$ to know about it. 

The absence of a smearing function makes the holographic dictionary more elusive. Bulk evolution combined with boundary evolution are not even in principle sufficient to answer the question of what the precursers are. Other methods must be developed to find the dictionary between  bulk operators and nonlocal boundary operators.

\acknowledgments
We are grateful to Raphael Bousso and Ben Freivogel for many discussions and insights. We thank Steve Giddings, Don Marolf, Lenny Susskind, Claire Zukowski and especially Douglas Stanford for helpful discussions.  The work of VR is supported by the Berkeley Center for Theoretical Physics. The work of SL is supported by a John A. McCone Postdoctoral Fellowship, the U.S. Department of Energy under contract No. DE-FG02-92ER40701, and the Gordon and Betty Moore Foundation through Grant No. 776 to the Caltech Moore Center for Theoretical Cosmology and Physics.

\bibliographystyle{utcaps}
\bibliography{all}

\end{document}